\documentclass[trackchanges]{aastex701}
\usepackage{amsmath}
\usepackage{graphicx}
\usepackage{subcaption} 
\usepackage{threeparttable}
\usepackage{multirow}
\usepackage{booktabs}

\begin{document}

\title{\emph{Gaia} DR3 Compact Binary Candidates Constrained by Radial Velocity Measurements}

\author[]{Ying-jie Li}
\affiliation{Purple Mountain Observatory, Chinese Academy of Sciences, Nanjing 210008,People’s Republic of China;}
\affiliation{State Key Laboratory of Radio Astronomy and Technology, Purple Mountain Observatory, Chinese Academy of Sciences, 10 Yuanhua Road, Nanjing 210023, China}
\affiliation{University of Science and Technology of China, Jinzhai Road, Hefei 230026, People’s Republic of China;}
\email[]{}

\author[]{Xianjin Shen}
\affiliation{Purple Mountain Observatory, Chinese Academy of Sciences, Nanjing 210008,People’s Republic of China;}
\affiliation{University of Science and Technology of China, Jinzhai Road, Hefei 230026, People’s Republic of China;}
\email[]{}

\author[]{Zehao Lin}
\affiliation{Purple Mountain Observatory, Chinese Academy of Sciences, Nanjing 210008,People’s Republic of China;}
\affiliation{State Key Laboratory of Radio Astronomy and Technology, Purple Mountain Observatory, Chinese Academy of Sciences, 10 Yuanhua Road, Nanjing 210023, China}
\email[]{}

\author[]{Jingjing Li}
\affiliation{Purple Mountain Observatory, Chinese Academy of Sciences, Nanjing 210008,People’s Republic of China;}
\affiliation{State Key Laboratory of Radio Astronomy and Technology, Purple Mountain Observatory, Chinese Academy of Sciences, 10 Yuanhua Road, Nanjing 210023, China}
\affiliation{University of Science and Technology of China, Jinzhai Road, Hefei 230026, People’s Republic of China;}
\email[]{}

\author[]{Dejian Liu}
\affiliation{College of Mathematics and Physics, China Three Gorges University, Yichang 443000, People’s Republic of China;}
\affiliation{Center for Astronomy and Space Sciences, China Three Gorges University, Yichang 443000, People’s Republic of China;}
\email[]{}

\author[]{Kangteng Chang}
\affiliation{College of Mathematics and Physics, China Three Gorges University, Yichang 443000, People’s Republic of China;}
\affiliation{Center for Astronomy and Space Sciences, China Three Gorges University, Yichang 443000, People’s Republic of China;}
\email[]{}

\author[]{Yao Huang}
\affiliation{College of Mathematics and Physics, China Three Gorges University, Yichang 443000, People’s Republic of China;}
\affiliation{Center for Astronomy and Space Sciences, China Three Gorges University, Yichang 443000, People’s Republic of China;}
\email[]{}

\author[]{Zhenyu Wu}
\affiliation{National Astronomical Observatories, Chinese Academy of Sciences, 20A Datun Road, Chaoyang District, Beijing 100101, People’s Republic of China;}
\affiliation{School of Astronomy and Space Science, University of Chinese Academy of Sciences, Beijing 101408, People’s Republic of China}
\email[]{}

\author[]{Jianrong Shi}
\affiliation{National Astronomical Observatories, Chinese Academy of Sciences, 20A Datun Road, Chaoyang District, Beijing 100101, People’s Republic of China;}
\email[]{}

\author[]{Ye Xu}
\altaffiliation{Corresponding Author}
\affiliation{Purple Mountain Observatory, Chinese Academy of Sciences, Nanjing 210008,People’s Republic of China;}
\affiliation{State Key Laboratory of Radio Astronomy and Technology, Purple Mountain Observatory, Chinese Academy of Sciences, 10 Yuanhua Road, Nanjing 210023, China}
\affiliation{University of Science and Technology of China, Jinzhai Road, Hefei 230026, People’s Republic of China;}
\affiliation{Xinjiang Astronomical Observatory, Chinese Academy of Sciences, Urumqi, Xinjiang, 830011, People's Republic of China}
\email[show]{xuye@pmo.ac.cn} 

\begin{abstract}
Astrometry enables the most direct determination of the orbital elements and component masses of compact binaries. The potential of the current \emph{Gaia} data for this purpose has not yet been fully explored. In this work, we combine \emph{Gaia} astrometric and radial velocity (RV) measurements to perform joint orbital fitting for 1,835 \emph{Gaia} astrometric binaries with available RV data. Among the 518 systems with at least three effective RV epochs, we identify 171 compact binary candidates, including four with inferred dark companion masses exceeding 2.1 $M_{\odot}$ and 22 between 1.4 and 2.1 $M_{\odot}$. These candidates provide valuable targets for future follow-up observations and \emph{Gaia} DR4 analyses. We also flag 1,317 additional potential compact binary candidates with effective RV epochs less than three, including 84 with inferred companion masses above 2.1 $M_{\odot}$ and 173 between 1.4 and 2.1 $M_{\odot}$. More importantly, our results point to a large, currently hidden population of black hole and neutron star binaries that can be further explored with the forthcoming \emph{Gaia} DR4 data and complementary follow-up observations.
\end{abstract}

\keywords{\uat{Astrometric binary stars}{79} --- \uat{Compact binary stars}{283} --- \uat{Compact objects}{288} --- \uat{Radial velocity}{1332}}


\section{Introduction} \label{sect:intro}

Compact objects provide a unique link between astronomy and fundamental physics, while compact binaries combine the scientific value of compact objects and binary systems \citep{Horch2013}. They serve as natural laboratories for probing nuclear and particle physics under extreme conditions \citep{Lorimer+2008}, testing relativistic effects \citep{Deller+2018}, and constraining the possible time derivative of the Newtonian gravitational constant \citep{Ding+2023}. They also provide valuable insights into the formation and evolution of compact objects and binaries under extreme physical conditions, as well as the role of star cluster and galactic environments in shaping their properties.

With the advent of the multi-messenger era, various techniques have been developed to measure compact object masses, including gravitational-wave observations \citep[mainly for merging compact binaries;][]{Abbott+2023}, joint analyses of eclipsing-binary light curves and radial velocities \citep[RV, for nearly edge-on systems;][]{Jayasinghe+2021}, pulsar timing \citep{Kramer+2006}, and astrometry. Among these methods, long-term astrometric monitoring of binary orbital motion provides one of the most direct and least model-dependent approaches to determining binary masses and orbital elements.

\emph{Gaia} optical astrometry enables the determination of the mass and orbital elements of an unseen companion by tracing the orbital motion of the optically visible star, provided that the mass of the visible star is known. This approach has been demonstrated by recent discoveries. For example, \citet{GaiaBH3+2024} combined multi-epoch \emph{Gaia} astrometry with RV measurements to determine the mass of the black hole (BH) in \emph{Gaia} BH3, yielding $32.70 \pm 0.82$ $M_{\odot}$, the most massive stellar-mass BH currently known in the Milky Way. Earlier joint \emph{Gaia} astrometry and RV analyses identified two BH binary candidates (\emph{Gaia} BH1, \citealt{GaiaBH1+2023}; \emph{Gaia} BH2, \citealt{GaiaBH2+2023}) and 21 neutron star (NS) binary candidates \citep{GaiaNS1+2024, GaiaNSothers+2024}. The peculiar orbital properties of these systems, including their unexpectedly long periods, have motivated further theoretical studies \citep{El-Badry_BH3_formation_2024, Pina+2025}.

At radio wavelengths, \citet{Miller-Jones+2021} used Very Long Baseline Interferometry (VLBI) to trace the orbital motion of the stellar-mass BH in X-ray binary Cygnus X-1 by monitoring the jet radio photosphere. Combined with optical light curves and radial velocity measurements of the optically visible companion star, these observations yielded a BH mass of $21.2 \pm 2.2$ $M_{\odot}$. VLBI has also been used to measure the orbital motion of the pulsar PSR B1259-63 \citep{Miller-Jones+2018}.

Optical and radio astrometry provide complementary views of compact binaries: \emph{Gaia} traces the orbital motion of the optically luminous component, whereas VLBI can track the motion of the compact object. However, compact binaries currently occupy largely distinct regions of orbital-period space, with little overlap between \emph{Gaia}-identified systems and previously known NS and BH binaries \citep{El-Badry2024, GaiaNSothers+2024}. Gaia DR3 remains primarily sensitive to long-period systems, while the forthcoming Gaia DR4 is expected to enable the detection of shorter-period compact binaries \citep{El-Badry+2024}. 
Meanwhile, SKA-VLBI, which incorporates the highly sensitive Square Kilometre Array (SKA) into a VLBI network, will enable the detection of fainter and more numerous compact binaries, including pulsar binaries, and the monitoring of their orbital motion \citep{Li+2024-SKA-VLBI, Lin+2026}.
The combination of radio and optical astrometry will open a new avenue for measuring both components' orbital motions simultaneously, providing largely model-independent constraints on compact binary masses and offering a geometric-dynamical benchmark for testing theories of stellar and binary evolution.

In this context, we perform a systematic search of the \emph{Gaia} DR3 Non-Single Star (NSS) catalog \citep{GaiaNSS+2023} for compact binary candidates. The NSS catalog contains $\sim1.7\times10^5$ astrometric binaries, whose photocenter motions provide information on the orbital motion induced by unseen companions, allowing their dynamical properties (e.g., companion masses) to be constrained. RV measurements provide complementary information by probing the line-of-sight orbital motion of the visible component. The combination of \emph{Gaia} astrometry and RV measurements has revealed several compact binary candidates, including BH and NS candidates.

However, the known population of compact binary candidates remains limited due to the long observational baselines required for RV measurements \citep[typically hundreds of days or longer; e.g.,][]{GaiaBH1+2023, GaiaBH2+2023, GaiaNS1+2024}. Motivated by this limitation, we combine the \emph{Gaia} DR3 NSS catalog with archival RV measurements from LAMOST \citep{Cui+2012,Luo+2015, Liu+2020}, APOGEE \citep{Meszaros+2025,SDSS+2025}, and GALAH \citep{Buder+2025} to search for compact binary candidates. Although forthcoming \emph{Gaia} DR4 will provide epoch RVs and astrometric transits, the DR3-based candidates identified here can provide valuable targets for ground-based multi-epoch follow-up observations \citep[e.g.,][]{Makarov-Tokovinin2024}. The resulting sample will provide valuable targets for future orbital characterization, dynamical mass measurements, and population studies. 

\section{Sample} \label{sec:data}

For each system in the NSS catalog, the photocenter orbital semi-major axis, $a_{0,\rm{ast}}$, is calculated following the method described by \citet{Halbwachs+2023}. Combining $a_{0,\rm{ast}}$ with the orbital period, $P_{\rm{ast}}$, and parallax, $\varpi$, provided by the NSS catalog, the astrometric mass function is then calculated as
\begin{equation}\label{equ:mass_funnction_ast}
    f_{\rm M, ast} = \left(\frac{a_{0,\rm{ast}}}{\varpi}\right)^{3}\left(\frac{1~{\rm{yr}}}{P_{\rm{ast}}}\right)^2 \; M_{\odot},
\end{equation}
where both $a_{0,\rm{ast}}$ and $\varpi$ are in units of mas. Assuming a dark companion and a known mass of the visible star $M_{1,\rm{ast}}$, the mass of the unseen companion, $M_{2,\rm{ast}}$, can be inferred from the mass function
\begin{equation}\label{equ:M2_ast}
    f_{\rm {M, ast}} = M_{2,\rm{ast}} \left(\frac{M_{2,\rm{ast}}}{M_{1,\rm{ast}} + M_{2,\rm{ast}}}\right)^{2}.
\end{equation}
The \emph{Gaia} \texttt{astrophysical\_parameters} catalog provides photometric mass estimates (\texttt{mass\_flame}) for $\sim 9\times10^4$ binary systems, which we adopt as $M_{1,\rm{ast}}$ to derive $M_{2,\rm{ast}}$. The reported metallicity uncertainties in the \emph{Gaia} \texttt{astrophysical\_parameters} catalog are known to be underestimated \citep{Brandner+2023}, and the \texttt{mass\_flame} estimates may be affected by systematic biases at certain evolutionary stages, such as the red giant branch \citep[e.g.,][]{Babusiaux+2023}. Therefore, we do not adopt the official uncertainties and instead conservatively assign a uniform 30\% uncertainty to all mass estimates.

The lowest-mass NS currently known has a mass of $\sim$0.77 $M_{\odot}$ \citep{Doroshenko+2022}. We therefore select binary systems with $M_{2,\rm{ast}} \geq 0.7$ $M_{\odot}$, yielding 17,191 sources. Among them, 425 have $M_{2,\rm{ast}} > 2.1$ $M_{\odot}$, 2,230 have $1.4 \leq M_{2,\rm{ast}} \leq 2.1$ $M_{\odot}$, and 14,536 have $0.7 \leq M_{2,\rm{ast}} < 1.4$ $M_{\odot}$. \citet{Liu+2026} has cross-matched the cluster member catalog of \citet{Hunt+2024} with the NSS catalog. To avoid duplication, we exclude the binaries identified in that work and focus on the remaining systems. Cross-matching with archival RV measurements from LAMOST \citep[including medium-resolution MRS and low-resolution LRS data;][]{Cui+2012,Luo+2015, Liu+2020}, APOGEE \citep{Meszaros+2025,SDSS+2025}, and GALAH \citep{Buder+2025} yields 1,835 sources with available RV measurements. Among these, 19 have $M_{2,\rm{ast}} > 2.1$ $M_{\odot}$, 216 have $1.4 \leq M_{2,\rm{ast}} \leq 2.1$ $M_{\odot}$, and 1,600 have $0.7 \leq M_{2,\rm{ast}} < 1.4$ $M_{\odot}$. The basic properties of the 1,835 sources, including $P_{\rm{ast}}$, eccentricity $e_{\rm{ast}}$, $a_{0,\rm{ast}}$, $M_{1,\rm{ast}}$, $M_{2,\rm{ast}}$, $f_{\rm M, ast}$, $\varpi$, \emph{Gaia} G-band apparent magnitude $G$, and NSS type, are listed in Table \ref{table1}. 

Because systematic offsets may exist between RV datasets, combining RV measurements involves accounting for these offsets, with LAMOST MRS and LRS treated as separate datasets. This may reduce the number of effective RV epochs. We therefore define an effective number of RV epochs, $n_{\rm eff}$ (see Appendix \ref{app:RV_epoch}), and list both the total number of RV epochs, $n_{\rm RV}$, and $n_{\rm eff}$ in Table \ref{table1}. Among the 1,835 sources, 146 have $n_{\rm eff}\geq 7$, including 1, 21, and 124 with $M_{2,\rm{ast}}>2.1$ $M_{\odot}$, $1.4\leq M_{2,\rm{ast}}\leq2.1$ $M_{\odot}$, and $0.7\leq M_{2,\rm{ast}}<1.4$ $M_{\odot}$, respectively. An additional 372 have $3\leq n_{\rm eff}<7$, including 6, 53, and 313 in the corresponding three mass ranges.

\begin{table}[htbp]
    \tiny
    \setlength\tabcolsep{3pt}
    \caption{Basic Properties of the \emph{Gaia} Binary Sample\label{table1}}
    \centering
    \begin{tabular}{lccccccccccc}
    \toprule
    Source ID & NSS type & $P_{\rm{ast}}$ & $e_{\rm{ast}}$ & $a_{\rm{0, ast}}$ & $M_{1,\rm{ast}}$ & $M_{2, \rm{ast}}$ & $f_{\rm M, ast}$ & $\varpi$ & $G$ & $n_{\rm{RV}}$ & $n_{\rm{eff}}$ \\ 
    & & [days] & & [mas] & [$M_{\odot}$] & [$M_{\odot}$] & [$M_{\odot}$] & [mas] & [mag] & \\
    \midrule
4491401821468887552 & OrbitalAlternative & 31.31 $\pm$ 0.02  & 0.24 $\pm$ 0.08  & 1.30 $\pm$ 0.05  & 0.7 $\pm$ 0.2  & 130.1 $\pm$ 24.3  & 128.51 $\pm$ 24.49  & 1.32 $\pm$ 0.08  & 16.49  & 1 & 1 \\
824344003814438528 & OrbitalAlternative & 31.50 $\pm$ 0.04  & 0.41 $\pm$ 0.07  & 0.98 $\pm$ 0.03  & 0.8 $\pm$ 0.2  & 68.3 $\pm$ 8.5  & 67.08 $\pm$ 8.70  & 1.24 $\pm$ 0.04  & 15.50  & 1 & 1 \\
3664684869697065984 & AstroSpectroSB1 & 1220.45 $\pm$ 233.17  & 0.46 $\pm$ 0.08  & 1.07 $\pm$ 0.21  & 3.0 $\pm$ 0.9  & 6.7 $\pm$ 3.7  & 3.20 $\pm$ 2.87  & 0.32 $\pm$ 0.05  & 11.56  & 4 & 4 \\
220123911510053248 & AstroSpectroSB1 & 758.76 $\pm$ 3.94  & 0.48 $\pm$ 0.02  & 0.92 $\pm$ 0.02  & 5.0 $\pm$ 1.5  & 3.2 $\pm$ 0.6  & 0.50 $\pm$ 0.05  & 0.71 $\pm$ 0.02  & 11.99  & 4 & 4 \\
1594082579405063168 & Orbital & 833.81 $\pm$ 11.13  & 0.62 $\pm$ 0.06  & 0.91 $\pm$ 0.05  & 2.3 $\pm$ 0.7  & 3.0 $\pm$ 0.5  & 1.00 $\pm$ 0.18  & 0.52 $\pm$ 0.01  & 12.28  & 3 & 3 \\
... \\
    \bottomrule
    \end{tabular}
    \begin{tablenotes}
        \item[] (This table is available in its entirety in machine-readable form.)
     \end{tablenotes}
\end{table}

\section{Joint \emph{Gaia} Astrometry and RV Orbital Fitting Results}\label{sec:gaia+RV}

We adopt the dynamical modeling approach used by \citet{GaiaBH1+2023, GaiaBH2+2023} and perform joint \emph{Gaia} astrometric and RV orbital fitting (hereafter, \emph{Gaia} + RV fitting) for all 1,835 sources. Since the individual \emph{Gaia} astrometric transits are not publicly available in DR3, we use the orbital parameters from the \emph{Gaia} NSS catalog, together with their reported uncertainties, to construct the \emph{Gaia} likelihood and combine it with the archival RV likelihood. Systematic RV offsets among different datasets are included as free parameters in the fitting. The MCMC sampling and posterior estimation are performed using the \texttt{emcee} package \citep{Foreman-Mackey+2013}.

\begin{figure}[htbp]
    \centering
    \begin{subfigure}[b]{0.45\linewidth}
        \centering
        \includegraphics[width=0.9\linewidth]{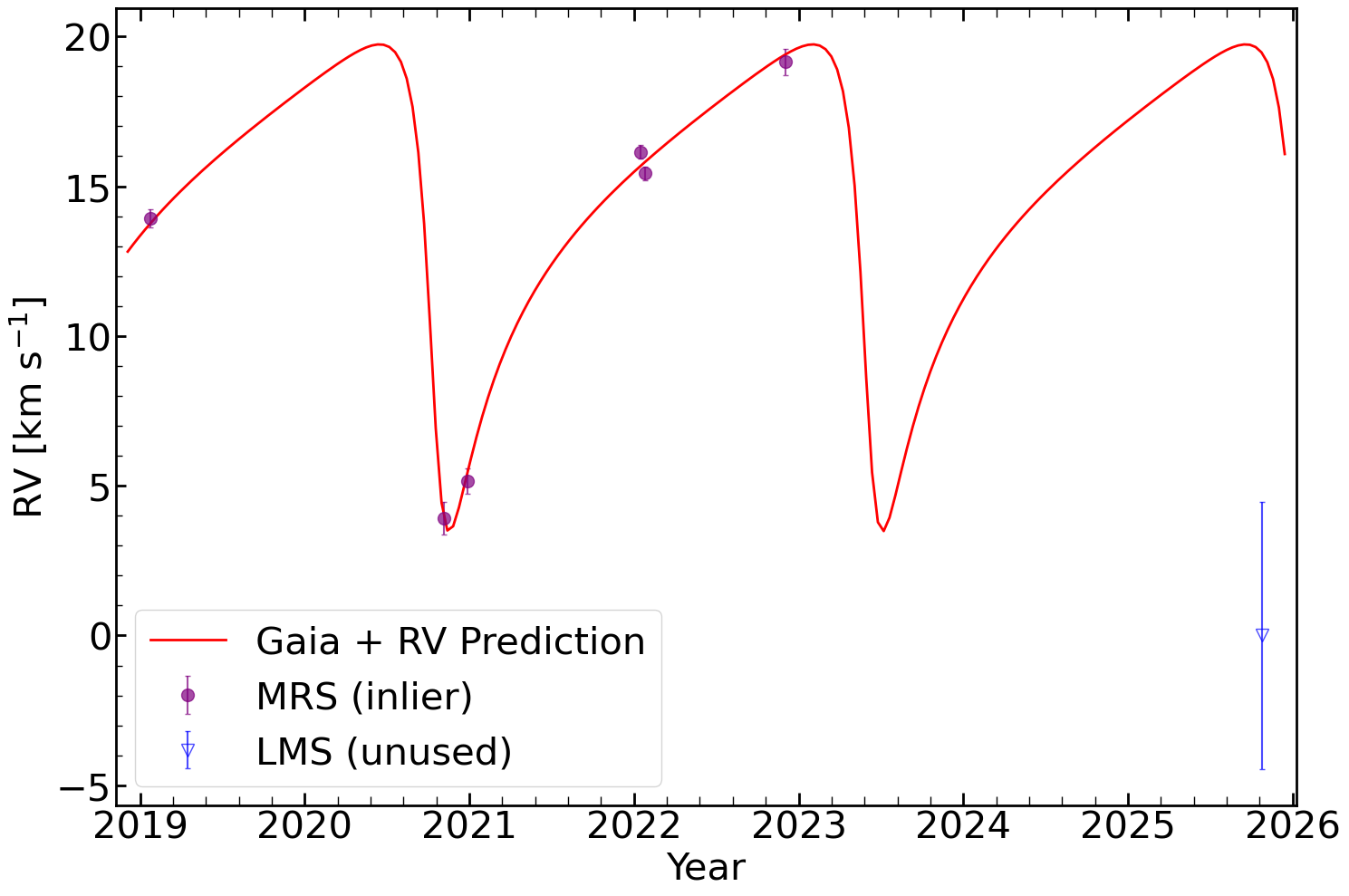}
        \includegraphics[width=0.9\linewidth]{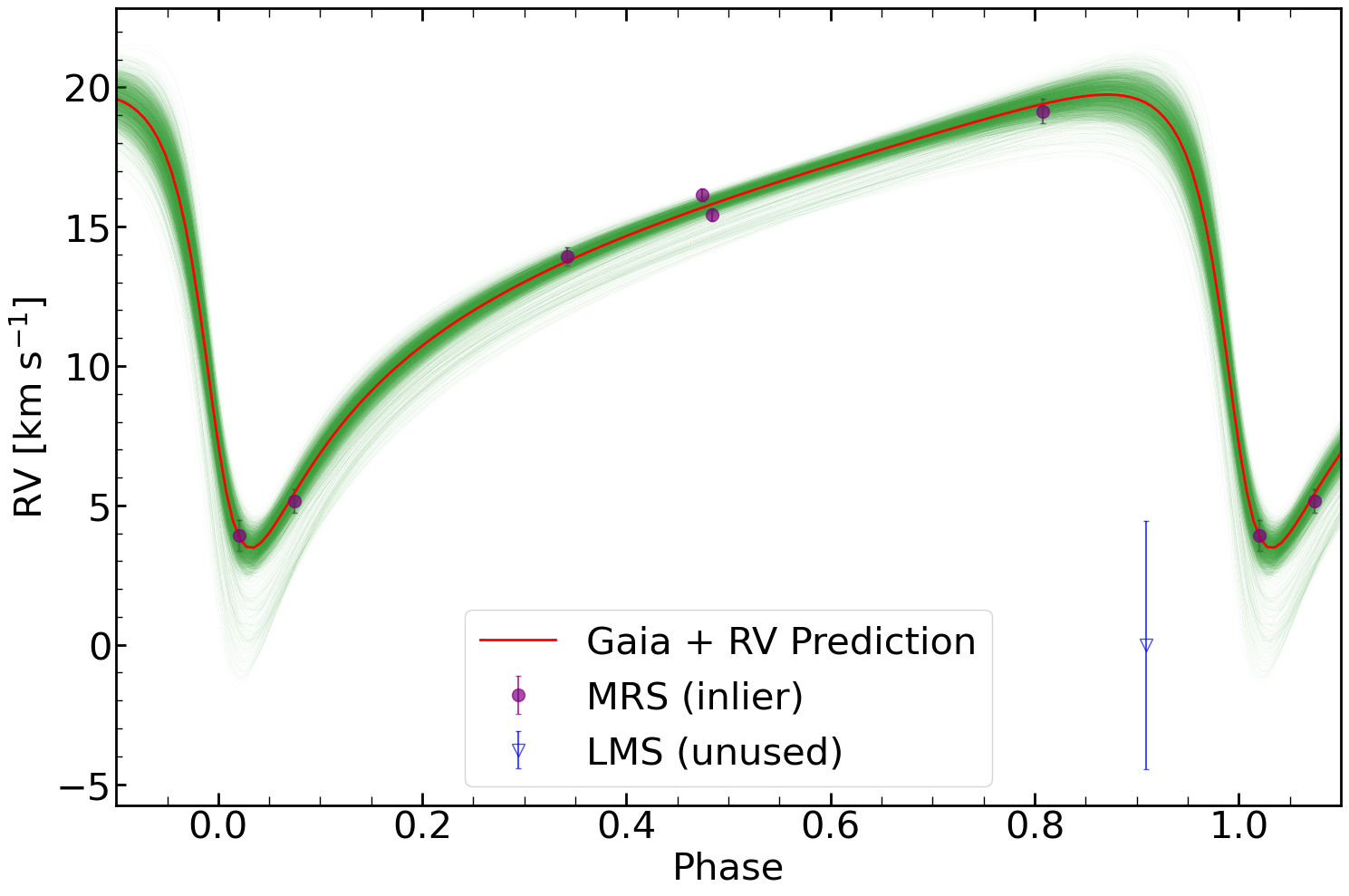} 
        \includegraphics[width=0.9\linewidth]{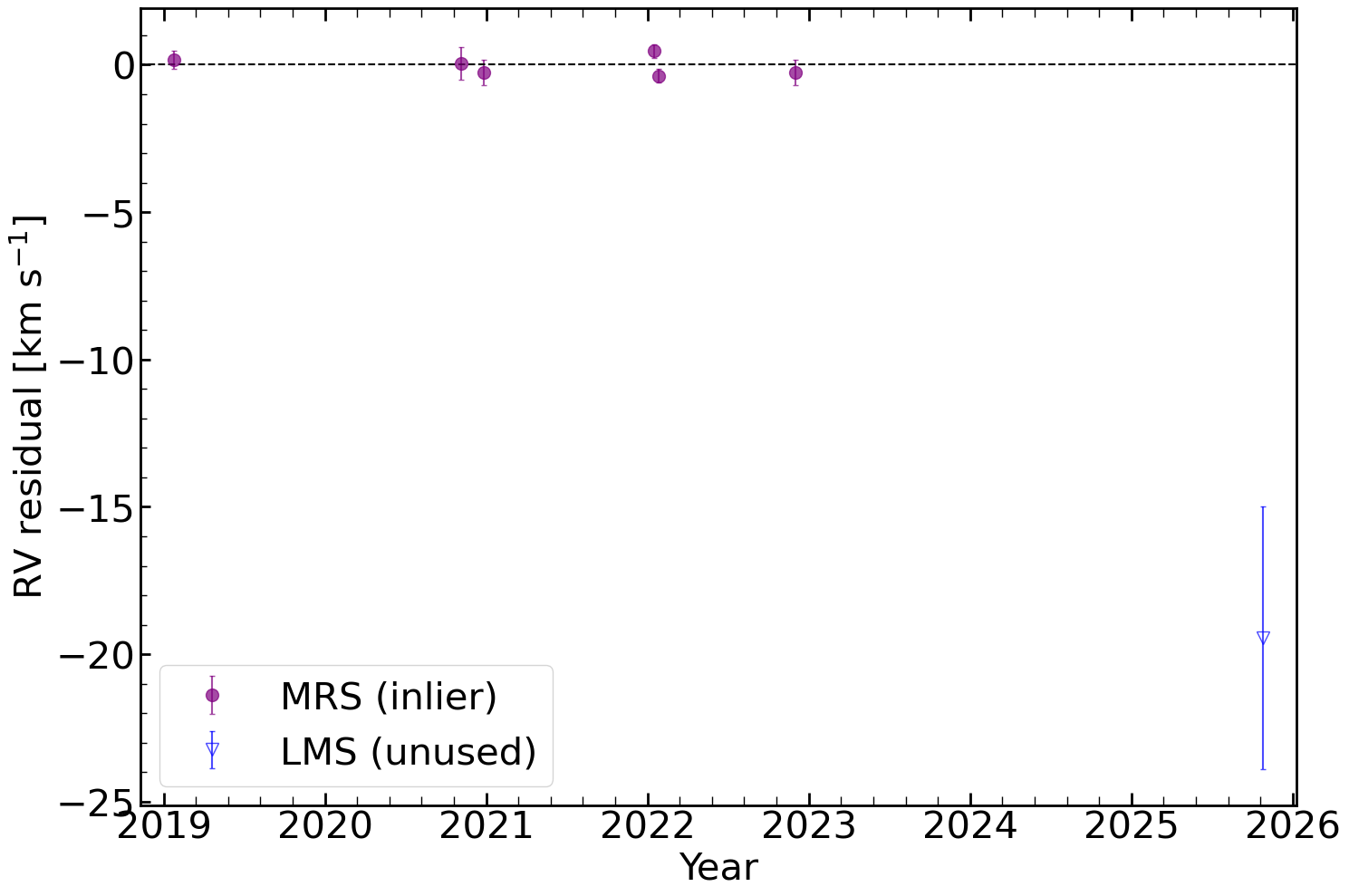}     
        \caption{\emph{Gaia} DR3 231878244865339264 (Good-Fit)}
    \end{subfigure}
    \begin{subfigure}[b]{0.45\linewidth}
        \centering
        \includegraphics[width=0.9\linewidth]{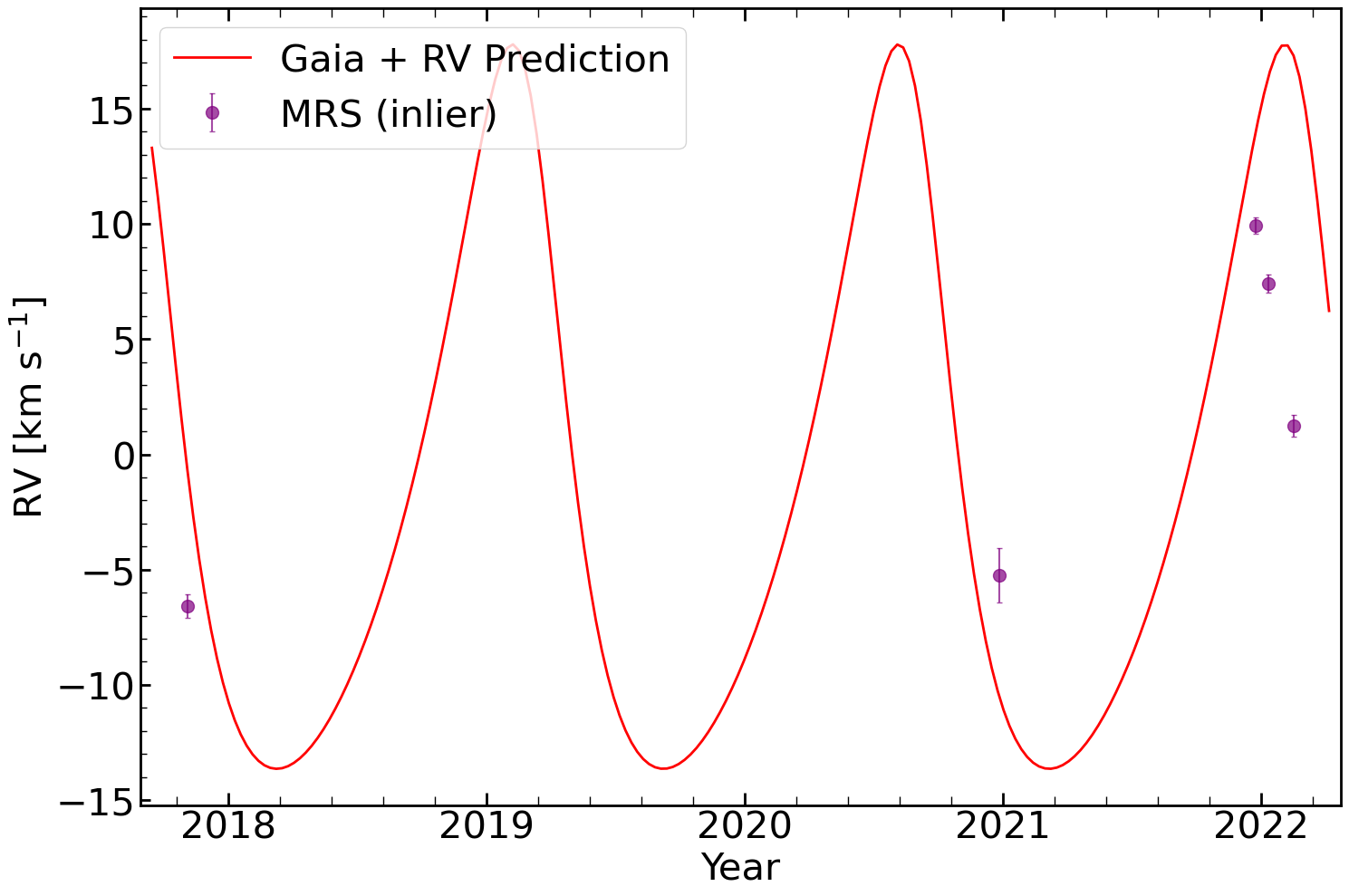}
        \includegraphics[width=0.9\linewidth]{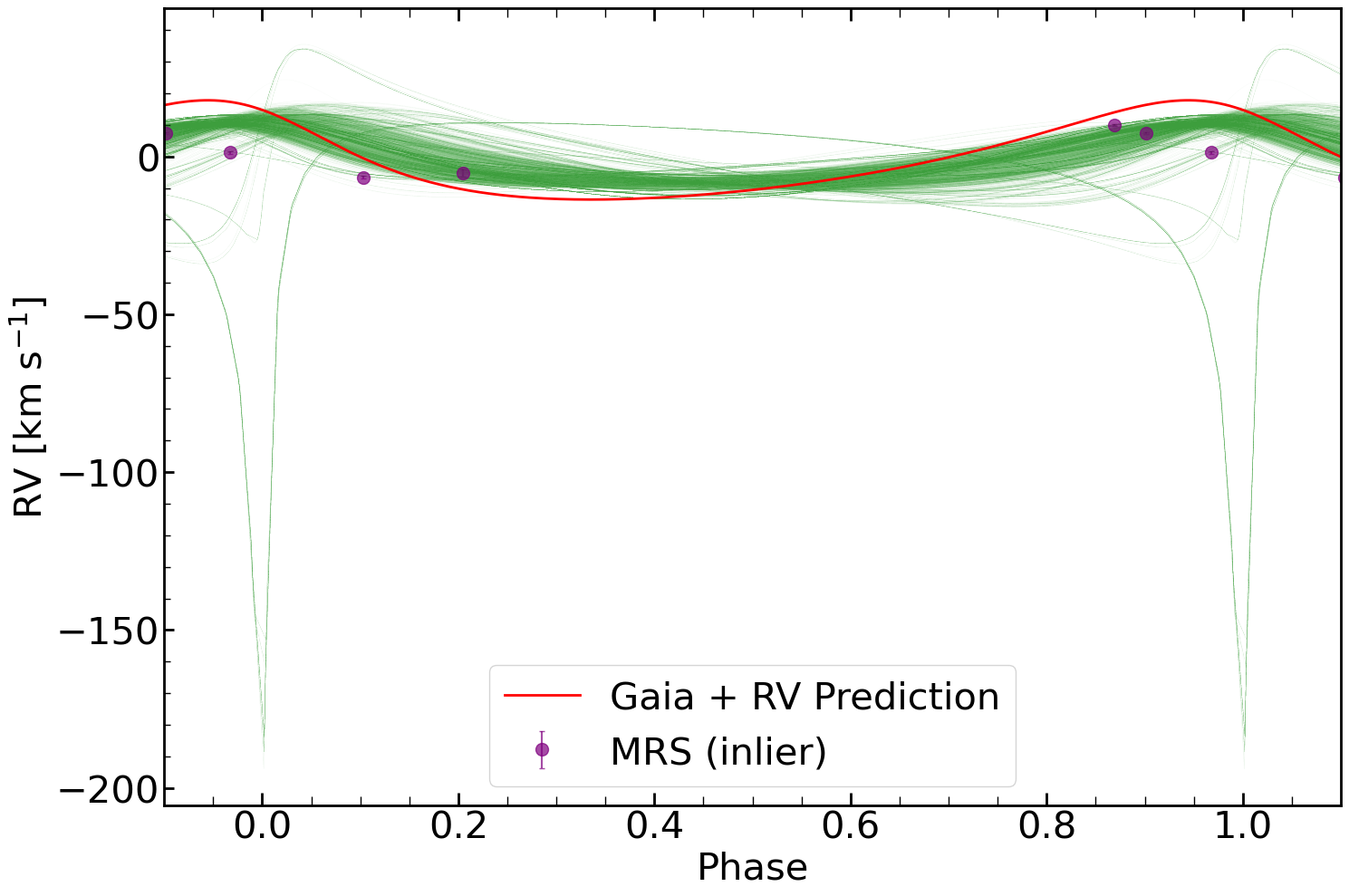} 
        \includegraphics[width=0.9\linewidth]{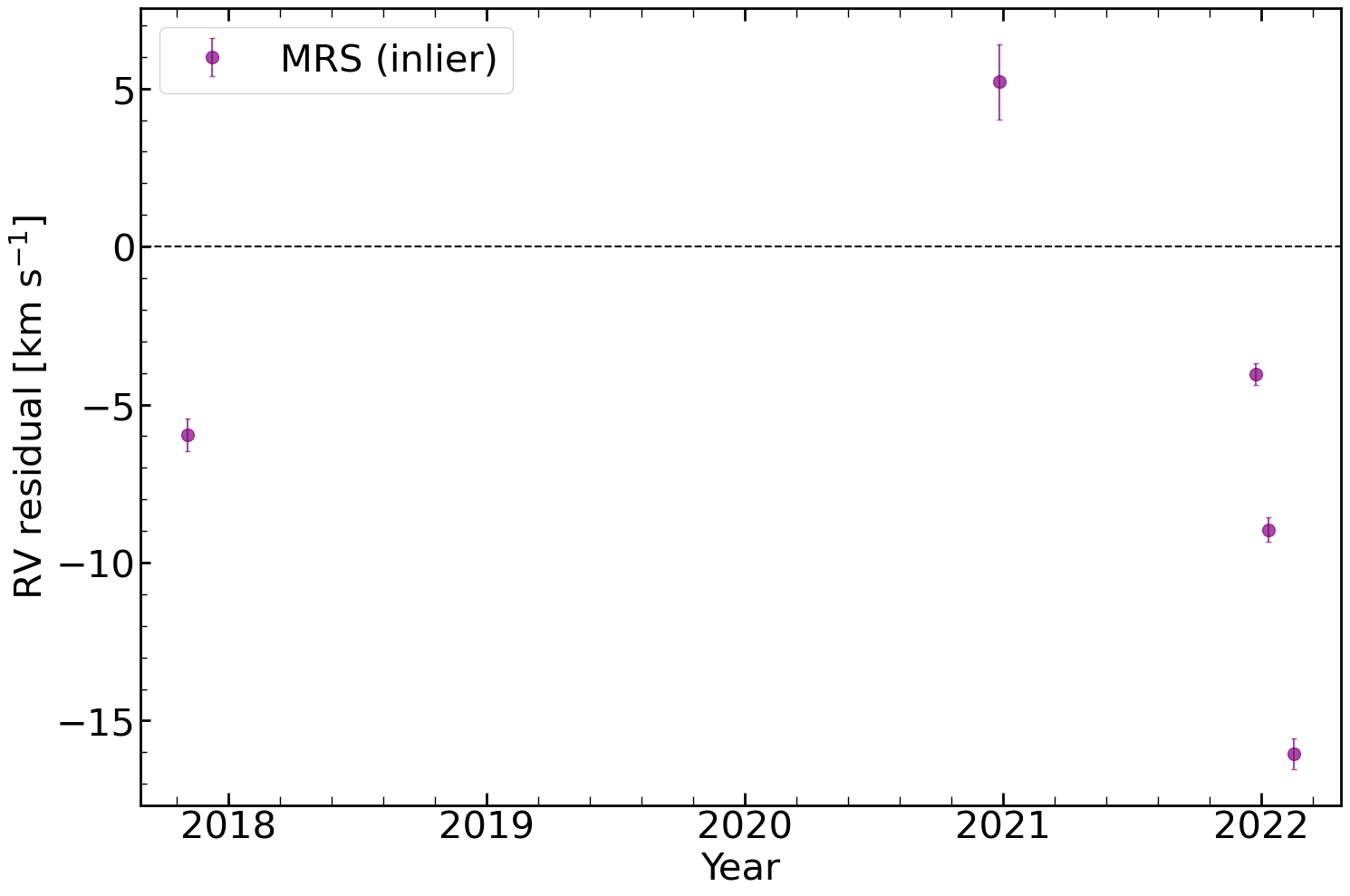}
        \caption{\emph{Gaia} DR3 157911142723659776 (Poor-Fit)}
    \end{subfigure}
    \caption{Examples of \emph{Gaia} + RV fitting results: (a) \emph{Gaia} DR3 231878244865339264 (Good-fit) and (b) 157911142723659776 (Poor-fit). From top to bottom: multi-epoch RV measurements, phase-folded RV curves, and residuals. Filled and open symbols denote inlier and unused/outlier RV measurements, respectively, with colors indicating different RV datasets. The red curve shows the median posterior model, while the light-green curves show randomly selected posterior samples. For weakly constrained fits, the median posterior model may deviate from the bulk of the posterior samples. In panel (a), the single LAMOST LRS measurement is excluded because only one epoch is available. The black dashed line marks zero residual.
\label{fig:example}}
\end{figure}

Figures \ref{fig:example} and \ref{fig:corner} show examples of the \emph{Gaia} + RV fitting results and the corresponding posterior distributions of model parameters, respectively. Table \ref{table2} lists the key fitted parameters, including the orbital period, $P_{\rm{fit}}$, eccentricity, $e_{\rm{fit}}$, inclination, $i_{\rm{fit}}$, argument of periastron, $\omega_{\rm{fit}}$, ascending node longitude, $\Omega_{\rm{fit}}$, periastron time, $T_{p,\rm{fit}}$, center-of-mass RV, $\gamma_{\rm{fit}}$, and component masses $M_{1,\rm{fit}}$ and $M_{2,\rm{fit}}$. The values of the goodness-of-fit criteria used for the Good/Poor classification are also provided in Table \ref{table2} (see Appendix~\ref{app:score}). Among the 1,835 sources, 518 have $n_{\rm eff}\geq3$, with 192 are classified as Good and 326 as Poor. The remaining 1,317 sources have $n_{\rm eff}<3$. For the full sample, 169 have $M_{2,\rm{fit}}>2.1$ $M_{\odot}$ and 260 have $1.4\leq M_{2,\rm{fit}}\leq2.1$ $M_{\odot}$ (see Table \ref{tab:summary} for a complete summary).

\begin{table}[htbp]
    \centering
    \tiny
    \setlength\tabcolsep{0.8pt}
    \caption{Results of Joint \emph{Gaia} Astrometric and RV Orbital Fitting\label{table2}}
    \begin{tabular}{ccccccccccccccccc}
    \toprule
    Source ID & $P_{\rm{fit}}$ & $e_{\rm{fit}}$ & $i_{\rm{fit}}$ & $\omega_{\rm{fit}}$ & $\Omega_{\rm{fit}}$ & $T_{p,\rm{fit}}$ & $\gamma_{\rm{fit}}$ & $M_{\rm{2, fit}}$ & $M_{1, \rm{fit}}$ & $f_{\rm{M, fit}}$ & $\chi^2$ & $z$ & $f_{2\sigma}$ & $n_{\rm{used}}$ & Pr$_{\rm{II}}$ & Pr$_{\rm{III}}$  \\ 
     & [days] & & [rad] & [rad] & [rad] & [days] & [km s$^{-1}$] & [$M_{\odot}$] & [$M_{\odot}$] & [$M_{\odot}$] & & & & & [\%] & [\%] \\ \midrule
\multicolumn{16}{l}{Good (192 Sources)} \\
\midrule
145545725720154240	&	710.49 	 $_{-5.01}^{+6.37}$ &	0.41 	 $_{-0.03}^{+0.04}$ &	1.37 	 $_{-0.05}^{+0.04}$ &	3.8 	 $_{-0.0}^{+0.1}$ &	1.9 	 $_{-0.0}^{+0.1}$ &	57026.0 	 $_{-13.2}^{+14.8}$ &	25.90 	 $_{-0.12}^{+0.12}$ &	3.2 	 $_{-0.2}^{+0.2}$ &	3.2 	 $_{-0.3}^{+0.3}$ &	0.80 	 $_{-0.06}^{+0.07}$ &	2.7 	&	1.3 	&	0.8 	&	4 	&		&		\\
420760358106619520	&	626.14 	 $_{-5.82}^{+4.57}$ &	0.56 	 $_{-0.06}^{+0.09}$ &	0.54 	 $_{-0.05}^{+0.09}$ &	2.0 	 $_{-0.1}^{+0.1}$ &	0.9 	 $_{-0.1}^{+0.1}$ &	57317.5 	 $_{-4.1}^{+5.2}$ &	-61.73 	 $_{-0.68}^{+0.53}$ &	2.8 	 $_{-0.2}^{+0.3}$ &	3.0 	 $_{-0.3}^{+0.4}$ &	0.66 	 $_{-0.06}^{+0.09}$ &	1.6 	&	0.5 	&	0.7 	&	3 	&		&		\\
809741149368202752	&	934.41 	 $_{-6.26}^{+5.62}$ &	0.43 	 $_{-0.12}^{+0.04}$ &	1.34 	 $_{-0.10}^{+0.04}$ &	0.6 	 $_{-0.2}^{+0.1}$ &	5.6 	 $_{-0.1}^{+0.6}$ &	57790.0 	 $_{-12.1}^{+12.1}$ &	-37.01 	 $_{-0.88}^{+0.38}$ &	2.3 	 $_{-0.5}^{+0.9}$ &	0.9 	 $_{-0.4}^{+0.5}$ &	1.15 	 $_{-0.17}^{+0.64}$ &	2.1 	&	1.0 	&	0.8 	&	6 	&	0.00 	&	100.00 	\\
212518967538774272	&	595.55 	 $_{-4.36}^{+2.99}$ &	0.39 	 $_{-0.03}^{+0.03}$ &	2.65 	 $_{-0.03}^{+0.03}$ &	6.1 	 $_{-0.1}^{+0.1}$ &	5.4 	 $_{-0.1}^{+0.1}$ &	57497.0 	 $_{-7.0}^{+6.3}$ &	18.37 	 $_{-0.30}^{+0.31}$ &	2.3 	 $_{-0.2}^{+0.2}$ &	2.6 	 $_{-0.4}^{+0.3}$ &	0.47 	 $_{-0.04}^{+0.06}$ &	2.8 	&	0.8 	&	0.8 	&	9 	&		&		\\

... \\
\midrule
\multicolumn{16}{l}{Poor (326 Sources)} \\
\midrule
3664684869697065984	&	1453.8 	 $_{-686.8}^{+662.6}$ &	0.58 	 $_{-0.25}^{+0.24}$ &	0.46 	 $_{-0.26}^{+0.24}$ &	5.5 	 $_{-2.0}^{+0.8}$ &	3.3 	 $_{-2.5}^{+2.0}$ &	57589.1 	 $_{-92.0}^{+61.6}$ &	-25.62 	 $_{-2.76}^{+4.85}$ &	18.1 	 $_{-16.3}^{+34.7}$ &	3.8 	 $_{-1.4}^{+0.9}$ &	12.17 	 $_{-11.69}^{+33.91}$ &	>100	&	>100	&	0.0 	&	4 	&		&		\\
3061746846610509824	&	1047.8 	 $_{-186.2}^{+167.0}$ &	0.12 	 $_{-0.05}^{+0.05}$ &	0.35 	 $_{-1.02}^{+0.05}$ &	4.4 	 $_{-3.7}^{+1.2}$ &	3.2 	 $_{-2.7}^{+2.3}$ &	57173.5 	 $_{-275.1}^{+177.0}$ &	66.64 	 $_{-0.45}^{+0.46}$ &	12.6 	 $_{-8.1}^{+9.0}$ &	2.9 	 $_{-0.6}^{+0.4}$ &	8.04 	 $_{-6.30}^{+8.53}$ &	>100	&	9.8 	&	0.0 	&	4 	&		&		\\
... \\
\midrule
\multicolumn{16}{l}{$n_{\rm{eff}} < 3$ (1317 Sources)}\\
\midrule
4491401821468887552	&	31.3 	 $_{-0.0}^{+0.0}$ &	0.30 	 $_{-0.09}^{+0.10}$ &	1.67 	 $_{-0.08}^{+0.10}$ &	0.1 	 $_{-0.5}^{+0.4}$ &	0.1 	 $_{-0.1}^{+0.1}$ &	57369.7 	 $_{-7.2}^{+16.3}$ &	178.50 	 $_{-5.60}^{+4.87}$ &	155.1 	 $_{-26.3}^{+73.0}$ &	0.8 	 $_{-0.3}^{+0.3}$ &	153.48 	 $_{-26.18}^{+72.96}$ &	>100	&	21.5 	&	0.0 	&	1 	&		&		\\
824344003814438528	&	31.5 	 $_{-0.1}^{+0.1}$ &	0.42 	 $_{-0.09}^{+0.08}$ &	1.44 	 $_{-0.11}^{+0.08}$ &	3.1 	 $_{-2.8}^{+0.4}$ &	0.2 	 $_{-3.1}^{+0.2}$ &	57376.4 	 $_{-3.4}^{+2.4}$ &	-97.96 	 $_{-6.19}^{+6.93}$ &	77.7 	 $_{-18.4}^{+43.7}$ &	0.7 	 $_{-0.3}^{+0.3}$ &	76.09 	 $_{-18.37}^{+44.02}$ &	19.1 	&	4.4 	&	0.0 	&	1 	&		&		\\
\bottomrule
    \end{tabular}
    \begin{tablenotes}
        \item[] (This table is available in its entirety in machine-readable form.)
     \end{tablenotes}
\end{table}

\begin{figure}[htbp]
    \centering
    \begin{subfigure}[b]{0.45\linewidth}
        \centering
        \includegraphics[width=\linewidth]{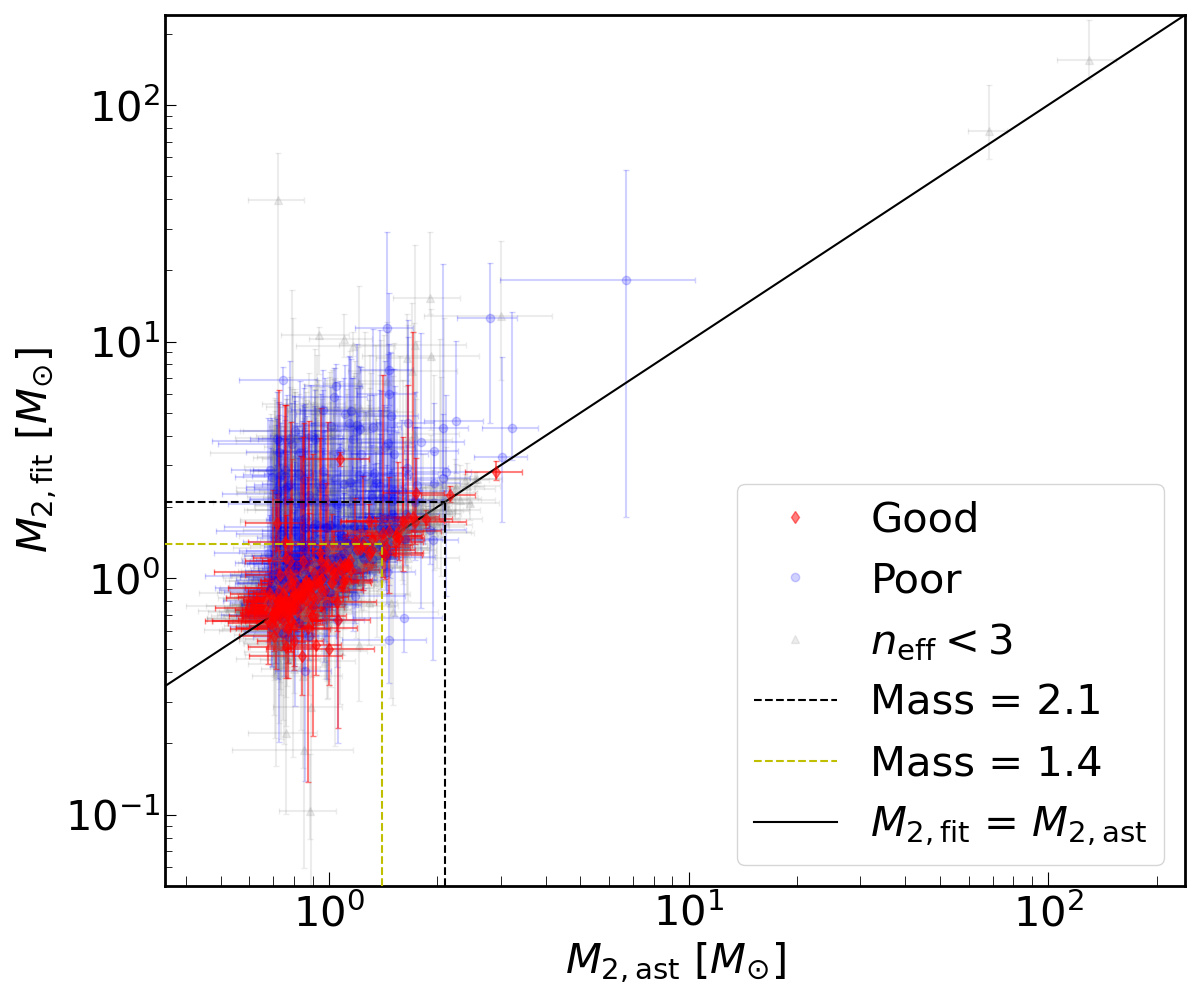}
        \caption{$M_{2,\rm{fit}}$ versus $M_{2,\rm{ast}}$}
    \end{subfigure}
    \begin{subfigure}[b]{0.45\linewidth}
        \centering
        \includegraphics[width=\linewidth]{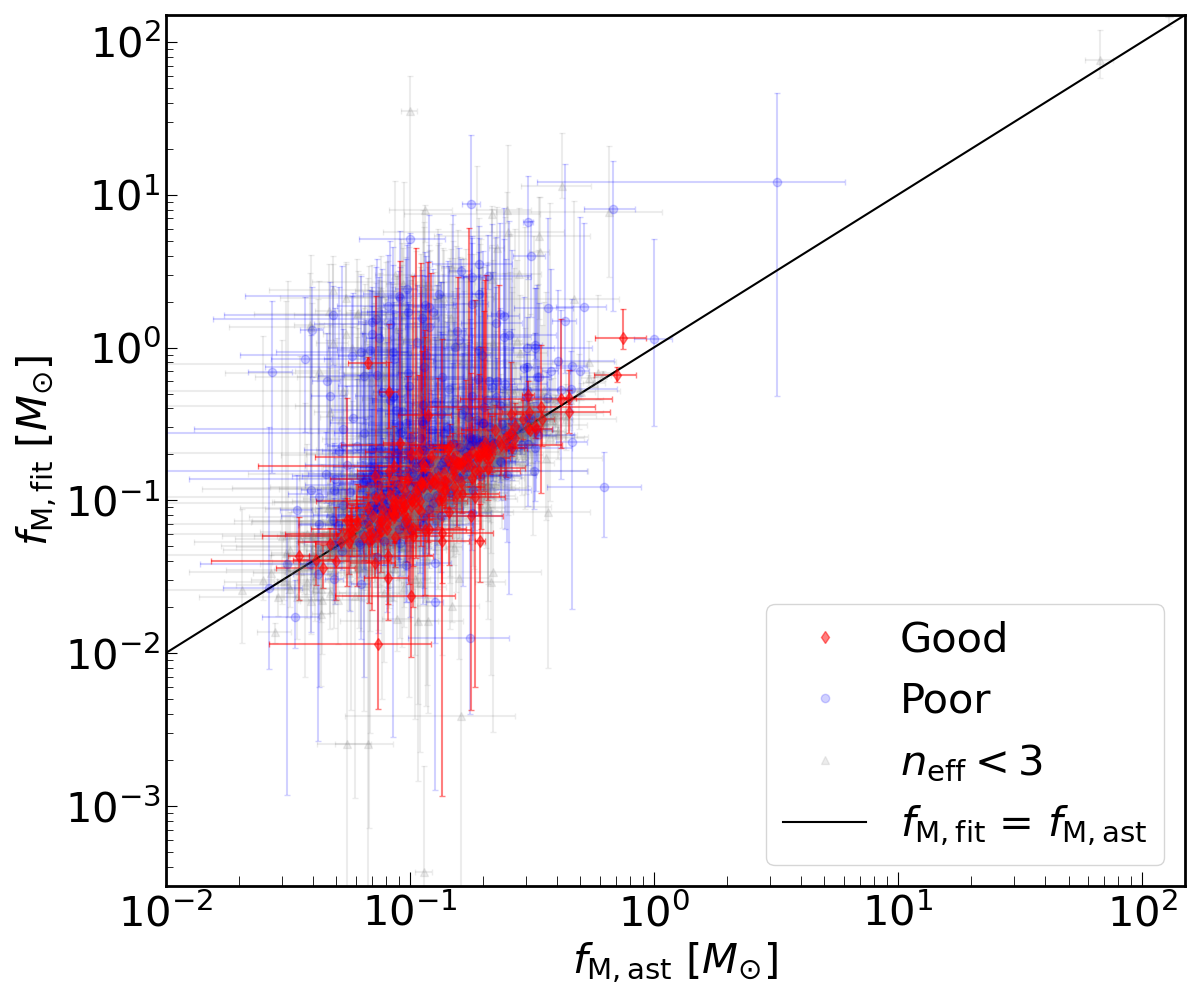}
        \caption{$f_{\rm{M,fit}}$ versus $f_{\rm{M,ast}}$}
    \end{subfigure}
    \caption{Comparison of the \emph{Gaia} + RV fitting results with the corresponding \emph{Gaia} astrometric estimates. (a) $M_{2,\rm{fit}}$ versus $M_{2,\rm{ast}}$.  The black solid line denotes $M_{2,\rm{fit}} = M_{2,\rm{ast}}$, while the black and yellow dashed lines correspond to masses of 2.1 and 1.4 $M_{\odot}$, respectively. (b) $f_{\rm{M,fit}}$ versus $f_{\rm{M,ast}}$, with the black solid line indicating $f_{\rm{M,fit}} = f_{\rm{M,ast}}$. Red, blue, and gray symbols represent Good-fit, Poor-fit, and $n_{\rm eff}<3$ sources, respectively.
    \label{fig:M2-astro-combine}}
\end{figure}

Figure \ref{fig:M2-astro-combine} compares $M_{2,\rm{fit}}$ with $M_{2,\rm{ast}}$ and $f_{\rm{M,fit}}$ with $f_{\rm{M,ast}}$. Both pairs show good agreement for Good-fit sources. Among the 192 Good-fit sources, four have $M_{2,\rm{fit}}>2.1$ $M_{\odot}$ (with a maximum mass of $\sim$3.2 $M_{\odot}$), two of which also have $M_{2,\rm{ast}}>2.1$ $M_{\odot}$. Among the same sample, 23 have $1.4\leq M_{2,\rm{fit}}\leq 2.1$ $M_{\odot}$, of which 16 fall within this range based on $M_{2,\rm{ast}}$ (see Table \ref{tab:summary}). Further statistical properties of these Good-fit sources, including the distributions of orbital and binary parameters and their correlations with $M_{2,\rm{fit}}$, are presented in Appendix \ref{app:statics}. Poor-fit sources and those with $n_{\rm eff}<3$ are not discussed further here.

\section{Candidate Classification and Analysis}\label{sec:discuss}

\citet{Shahaf+2019, Shahaf+2023} introduced the Astrometric Mass-Ratio Function (AMRF), a triage technique that uses \emph{Gaia} astrometric information to identify potential compact companions. Based on this method, \citet{Shahaf+2023} classified \emph{Gaia} binaries into three categories: Class (Cl) I, where the companion is most likely a single main-sequence star, although a close binary or a compact object cannot be excluded; Cl II, where the companion cannot be a single main-sequence star but can be either a close main-sequence binary or a compact object; and Cl III, where the companion cannot be either a single main-sequence star or a close main-sequence binary, making these systems the most promising compact object candidates. The classification in \citet{Shahaf+2023} does not cover all NSS sources and focuses on systems with a visible-star mass of $\lesssim$2 $M_{\odot}$.

We cross-matched our sample with the latest AMRF classifications from \citet{Shahaf+2023} using \emph{Gaia} IDs. This yielded 663 sources with AMRF classifications, including 73 Good-fit sources. Below, we focus on the Good-fit sources, with a summary provided in Table~\ref{tab:summary}. Sources with $\rm{Pr}_{\rm III}>50\%$, $\rm{Pr}_{\rm II}>50\%$, no AMRF classification, and $\rm{Pr}_{\rm I}>50\%$ are classified as Critical, High-priority, Medium-priority, and Low-priority compact binary candidates, respectively, where $\rm{Pr}_{\rm I}$, $\rm{Pr}_{\rm II}$, and $\rm{Pr}_{\rm III}$ represent the probabilities of AMRF classes I, II, and III. Table \ref{table2} lists the corresponding $\rm{Pr}_{\rm II}$ and $\rm{Pr}_{\rm III}$ values for all sources if they are available.

Four sources have $M_{\rm{2, fit}} > 2.1$ $M_{\odot}$. One source, \emph{Gaia} DR3 809741149368202752 has $M_{\rm{2, fit}} = 2.3^{+0.9}_{-0.5}$ $M_{\odot}$, $n_{\rm eff}=6$, and is classified as Cl III with $\rm{Pr}_{\rm III}\sim100.00\%$, making it the Critical compact binary candidate. This source was also identified as a compact binary candidate by \citet{Shahaf+2023} and \citet{GaiaNSothers+2024}. The other three sources, \emph{Gaia} DR3 145545725720154240 ($M_{\rm{2, fit}}=3.2^{+0.2}_{-0.2}$ $M_{\odot}$), 420760358106619520 ($M_{\rm{2, fit}}=2.8^{+0.3}_{-0.2}$ $M_{\odot}$) and 212518967538774272 ($M_{\rm{2, fit}}=2.3^{+0.2}_{-0.2}$ $M_{\odot}$), lack AMRF classifications and are therefore categorized as Medium-priority candidates. If confirmed as compact binaries, these systems would host either BHs or unusually massive NSs.

Twenty-three sources have $1.4 \leq M_{\rm{2, fit}} \leq 2.1$ $M_{\odot}$. Among these, \emph{Gaia} DR3 2069577481034003456 and 1367630085126769408 have companion masses of $M_{\rm{2, fit}}=1.5^{+0.2}_{-0.3}$ and $1.4^{+0.2}_{-0.1}$ $M_{\odot}$, with $n_{\rm eff}=4$ and 7, respectively. Both are classified as Cl II with $\rm{Pr}_{\rm II}\sim100.00\%$, making them High-priority NS binary candidates. \emph{Gaia} DR3 51004287124853248 has $M_{\rm{2, fit}}=1.7^{+0.6}_{-0.4}$ $M_{\odot}$ and $n_{\rm eff}=7$, but its $\rm{Pr}_{\rm II}$ and $\rm{Pr}_{\rm III}$ are only 43.06\% and 0.25\%, respectively, and it is therefore classified as a Low-priority candidate. The remaining 20 sources lack AMRF classifications and are categorized as Medium-priority NS binary candidates. It is worth noting that low inferred companion masses do not necessarily rule out BH companions. Among the 518 sources with $n_{\rm eff}\geq3$, 26 with $M_{\rm{2, fit}} \geq 1.4$ $M_{\odot}$ are classified as Medium-priority or higher BH or NS binary candidates, corresponding to a fraction of $\sim$5\%. Assuming that this fraction applies to the parent sample of 17,191 sources with $M_{\rm{2,ast}} \geq 0.7$ $M_{\odot}$, the expected number of BH and NS binary candidates would be $\sim17,191\times5\%\sim860$. This suggests that a substantial population of compact binary candidates may remain hidden, particularly among sources lacking RV measurements.

152 sources have $0.7\leq M_{\rm{2,fit}}< 1.4$ $M_{\odot}$. One source, \emph{Gaia} DR3 976146530514822656, has $M_{\rm{2,fit}} = 1.1^{+0.2}_{-0.2}$ $M_{\odot}$, $n_{\rm eff} = 4$, and a Cl III classification ($\rm{Pr}_{\rm III}\approx96.54\%$), making it a Critical compact binary candidate. Among the remaining 151 sources, 46 are classified as High-priority candidates ($\rm{Pr}_{\rm II}>50\%$), 86 as Medium-priority candidates owing to the lack of AMRF classifications, and the remaining 19 as Low-priority candidates ($\rm{Pr}_{\rm I}>50\%$). These candidates may host NS or white dwarf (WD) companions.

Thirteen sources have $M_{\rm{2,fit}}<0.7$ $M_{\odot}$. Two are classified as High-priority compact binary candidates, ten as Medium-priority because they lack AMRF classifications, and one as Low-priority. These candidates may host WD companions.

\begin{table}[htbp]
\centering
\tiny
\setlength\tabcolsep{3pt}
\caption{Summary of \emph{Gaia} Astrometric and \emph{Gaia} + RV Fitting Results\label{tab:summary}}
\begin{tabular}{l c c c c c c c c c}
\toprule
\multicolumn{10}{c}{\textbf{\emph{Gaia} Astrometric Results}} \\
\midrule
 & $M_{2,\rm{ast}} > 2.1$ $M_{\odot}$ & & $1.4 \leq M_{2,\rm{ast}} \leq 2.1$ $M_{\odot}$ & & $0.7 \leq M_{2,\rm{ast}} < 1.4$ $M_{\odot}$ & & $M_{2,\rm{ast}} < 0.7$ $M_{\odot}$ & & Total \\
\midrule
$n_{\rm{eff}} <3$ & 12 & & 142 & & 1163 & & Not Selected & & 1317 \\
$3 \leq n_{\rm{eff}} < 7$ & 6 & & 53 & & 313 & & Not Selected & & 372 \\
$n_{\rm{eff}} \geq 7$ & 1 & & 21 & & 124 & & Not Selected & & 146 \\
Total & 19 & & 216 & & 1600 & & Not Selected & & 1835 \\
\midrule
\multicolumn{10}{c}{\textbf{\emph{Gaia} + RV Fitting Results}} \\
\midrule
 & $M_{2,\rm{fit}} > 2.1$ $M_{\odot}$ & & $1.4 \leq M_{2,\rm{fit}} \leq 2.1$ $M_{\odot}$ & & $0.7 \leq M_{2,\rm{fit}} < 1.4$ $M_{\odot}$ & & $M_{2,\rm{fit}} < 0.7$ $M_{\odot}$ & & Total \\
\midrule
Good & 4 & & 23 & & 152 & & 13 & & 192 \\
Poor & 81 & & 64 & & 172 & & 9 & & 326 \\
$n_{\rm{eff}} <3$ & 84 & & 173 & & 996 & & 64 & & 1317 \\
Total & 169 & & 260 & & 1320 & & 86 & & 1835 \\
\midrule
\multicolumn{10}{c}{\textbf{Good \emph{Gaia} + RV Fits}} \\
\midrule
 & \multicolumn{2}{c}{$M_{2,\rm{fit}} > 2.1$ $M_{\odot}$} & \multicolumn{2}{c}{$1.4 \leq M_{2,\rm{fit}} \leq 2.1$ $M_{\odot}$} & \multicolumn{2}{c}{$0.7 \leq M_{2,\rm{fit}} < 1.4$ $M_{\odot}$} & \multicolumn{2}{c}{$M_{2,\rm{fit}} < 0.7$ $M_{\odot}$} & Total \\
\cmidrule(lr){2-3} \cmidrule(lr){4-5} \cmidrule(lr){6-7} \cmidrule(lr){8-9}
 & $3 \leq n_{\rm{eff}} < 7$ & $n_{\rm{eff}} \geq 7$ & $3 \leq n_{\rm{eff}} < 7$ & $n_{\rm{eff}} \geq 7$ & $3 \leq n_{\rm{eff}} < 7$ & $n_{\rm{eff}} \geq 7$ & $3 \leq n_{\rm{eff}} < 7$ & $n_{\rm{eff}} \geq 7$ & \\
\midrule
$M_{2,\rm{ast}} > 2.1$ $M_{\odot}$ & 1 & 1 & 0 & 0 & 0 & 0 & 0 & 0 & 2 \\
$1.4 \leq M_{2,\rm{ast}} \leq 2.1$ $M_{\odot}$ & 1 & 0 & 11 & 5 & 1 & 1 & 0 & 0 & 19 \\
$M_{2,\rm{ast}} < 1.4$ $M_{\odot}$ & 1 & 0 & 5 & 2 & 123 & 27 & 8 & 5 & 171 \\
Total & 3 & 1 & 16 & 7 & 124 & 28 & 8 & 5 & 192 \\
\midrule
\multicolumn{10}{c}{\textbf{Classifications of Compact Binary Candidates with Good \emph{Gaia} + RV Fits}} \\ 
\midrule
 & \multicolumn{2}{c}{$M_{2,\rm{fit}} > 2.1$ $M_{\odot}$} & \multicolumn{2}{c}{$1.4 \leq M_{2,\rm{fit}} \leq 2.1$ $M_{\odot}$} & \multicolumn{2}{c}{$0.7 \leq M_{2,\rm{fit}} < 1.4$ $M_{\odot}$} & \multicolumn{2}{c}{$M_{2,\rm{fit}} < 0.7$ $M_{\odot}$} & Total \\
\cmidrule(lr){2-3} \cmidrule(lr){4-5} \cmidrule(lr){6-7} \cmidrule(lr){8-9}
 & $3 \leq n_{\rm{eff}} < 7$ & $n_{\rm{eff}} \geq 7$ & $3 \leq n_{\rm{eff}} < 7$ & $n_{\rm{eff}} \geq 7$ & $3 \leq n_{\rm{eff}} < 7$ & $n_{\rm{eff}} \geq 7$ & $3 \leq n_{\rm{eff}} < 7$ & $n_{\rm{eff}} \geq 7$ & \\
\midrule
Critical & 1 & 0 & 0 & 0 & 1 & 0 & 0 & 0 & 2 \\
High-priority & 0 & 0 & 1 & 1 & 40 & 6 & 1 & 1 & 50 \\
Medium-priority & 2 & 1 & 15 & 5 & 69 & 17 & 7 & 3 & 119 \\
Low-priority & 0 & 0 & 0 & 1 & 14 & 5 & 0 & 1 & 21 \\
Total & 3 & 1 & 16 & 7 & 124 & 28 & 8 & 5 & 192 \\
\bottomrule
\end{tabular}
\end{table}

Sources with $n_{\rm eff}<3$ (Table \ref{table2}) have limited RV measurements but still provide some constraints and may represent potential compact binary candidates. These 1,317 sources are distributed as follows: 84 have $M_{2,\rm fit}>2.1$ $M_{\odot}$, 173 have $1.4\leq M_{2,\rm fit}\leq2.1$ $M_{\odot}$, 996 have $0.7\leq M_{2,\rm fit}<1.4$ $M_{\odot}$, and 64 have $M_{2,\rm fit}<0.7$ $M_{\odot}$ (see Table \ref{tab:summary}). For Poor-fit sources (Table \ref{table2}), the corresponding counts in these four mass ranges are 81, 64, 172, and 9, respectively (see Table \ref{tab:summary}).


\section{Conclusions}

Since the release of the \emph{Gaia} DR3 NSS catalog, several studies have combined \emph{Gaia} astrometry with RV measurements to search for compact binaries. However, the number of identified BH and NS binary candidates remains limited, and the potential of \emph{Gaia} astrometry for systematic compact binary searches remains largely unexplored. 
In this work, we perform joint \emph{Gaia} astrometric and RV fitting for 1,835 \emph{Gaia} astrometric binaries using \emph{Gaia} astrometry and archival RV measurements from LAMOST (including MRS and LRS data), APOGEE, and GALAH. For the 518 sources with $n_{\rm eff}\geq3$, 192 are classified as Good-fit sources, including 171 Medium-priority or higher compact binary candidates. The main conclusions are summarized as follows:
\begin{enumerate}
\item One Critical compact binary candidate, \emph{Gaia} DR3 809741149368202752, and three Medium-priority candidates, 145545725720154240, 420760358106619520 and 212518967538774272, have $M_{2,\rm{fit}}>2.1$ $M_{\odot}$, with inferred companion masses of $\sim$2.3, $\sim$3.2, $\sim$2.8, and $\sim$2.3 $M_{\odot}$, respectively. These systems may host BH companions or massive NSs.
\item Twenty-two Medium-priority or higher compact binary candidates have $1.4\leq M_{2,\rm fit}\leq2.1$ $M_{\odot}$ and are likely to be NS binaries. The remaining 145 candidates with $M_{2,\rm fit}<1.4,M_{\odot}$ may correspond to NS or WD binaries. The possibility of BH companions cannot be ruled out in either group.
\item The 1,317 sources with $n_{\rm eff}<3$ remain potential compact binary candidates requiring further RV observations. Among them, 84 have $M_{2,\rm fit}>2.1$ $M_{\odot}$, and 173 have $1.4\leq M_{2,\rm fit}\leq2.1$ $M_{\odot}$.
\end{enumerate}
This work significantly expands the sample of potential compact binary candidates. A substantial population of BH and NS binaries may remain hidden among systems with limited or no RV measurements. The candidates identified here provide targets for future \emph{Gaia} DR4 and complementary ground-based follow-up observations. Such observations will provide critical opportunities to identify, confirm, and characterize these systems, opening a new window into compact binary populations.

\begin{acknowledgments}
We would like to thank the referee for reviewing the paper carefully and for the constructive comments that improved the paper. This work was funded by the National Key R\&D Program of China (grant No.2024YFA1611504); the NSFC Grands Nos. 12503071, 12403077, and 12403041; the National SKA Program of China (grant No. 2022SKA0120103); the Xinjiang Talent Development Fund (No. XJRC-2025-KJ-YJ-CXPT-180); and the Key Laboratory for Radio Astronomy. 

This work has made use of data from the European Space Agency (ESA) mission \emph{Gaia} (\url{www.cosmos.esa.int/gaia}), processed by the \emph{Gaia} Data Processing and Analysis Consortium (DPAC, \url{www.cosmos.esa.int/web/gaia/dpac/consortium}). 

This work made use of the data from LAMOST (Large Sky Area Multi-Object Fiber Spectroscopic Telescope, also known as the Guoshoujing Telescope) (\url{https://cstr.cn/31118.02.LAMOST}). LAMOST is a Chinese national mega-science facility, operated by National Astronomical Observatories, Chinese Academy of Sciences.

SDSS is managed by the Astrophysical Research Consortium for the Participating Institutions of the SDSS Collaboration, including Caltech, The Carnegie Institution for Science, Chilean National Time Allocation Committee (CNTAC) ratified researchers, The Flatiron Institute, the Gotham Participation Group, Harvard University, Heidelberg University, The Johns Hopkins University, L'Ecole polytechnique f{\' e}d{\' e}rale de Lausanne (EPFL), Leibniz-Institut f{$\rm \ddot u$}r Astrophysik Potsdam (AIP), Max-Planck-Institut f{$\rm \ddot u$}r Astronomie (MPIA Heidelberg), Max-Planck-Institut f{$\rm \ddot u$}r Extraterrestrische Physik (MPE), Nanjing University, National Astronomical Observatories of China (NAOC), New Mexico State University, The Ohio State University, Pennsylvania State University, Smithsonian Astrophysical Observatory, Space Telescope Science Institute (STScI), the Stellar Astrophysics Participation Group, Universidad Nacional Aut{\' o}noma de M{\' e}xico, University of Arizona, University of Colorado Boulder, University of Illinois at Urbana-Champaign, University of Toronto, University of Utah, University of Virginia, Yale University, and Yunnan University.

This work made use of the Fourth Data Release of the GALAH Survey \citep{Buder+2025}. The GALAH Survey is based on data acquired through the Australian Astronomical Observatory, under programs: A/2013B/13 (The GALAH pilot survey); A/2014A/25, A/2015A/19, A2017A/18 (The GALAH survey phase 1); A2018A/18 (Open clusters with HERMES); A2019A/1 (Hierarchical star formation in Ori OB1); A2019A/15, A/2020B/23, R/2022B/5, R/2023A/4, R2023B/5 (The GALAH survey phase 2); A/2015B/19, A/2016A/22, A/2016B/10, A/2017B/16, A/2018B/15 (The HERMES-TESS program); A/2015A/3, A/2015B/1, A/2015B/19, A/2016A/22, A/2016B/12, A/2017A/14, A/2020B/14 (The HERMES K2-follow-up program); R/2022B/02 and A/2023A/09 (Combining asteroseismology and spectroscopy in K2); A/2023A/8 (Resolving the chemical fingerprints of Milky Way mergers); and A/2023B/4 (s-process variations in southern globular clusters). 
\end{acknowledgments}

\appendix

\section{Overview of RV Epochs}\label{app:RV_epoch}

For the \emph{Gaia} + RV fitting, we define three RV epoch counts: the total number of RV epochs ($n_{\rm RV}$), the effective number of epochs ($n_{\rm eff}$), and the number of epochs used in the final fit ($n_{\rm used}$). Specifically, $n_{\rm RV}$ represents the total RV measurements collected from LAMOST (including MRS and LRS), APOGEE, and GALAH (Table \ref{table1}).

The $n_{\rm eff}$ is defined as follows. The RV dataset with the largest number of epochs is adopted as the reference dataset; in case of a tie, priority is assigned in the order of MRS, APOGEE, GALAH, and LRS, with MRS ranked first because it contributes the largest number of RV measurements among the datasets used in this work. For each non-reference dataset, one epoch is subtracted from its contribution; datasets with zero remaining epochs are marked as “unused”. The resulting total number of epochs is defined as $n_{\rm eff}$. After the initial \emph{Gaia} + RV fitting, clear outliers are removed and $n_{\rm eff}$ is updated accordingly (Table \ref{table1}). An outlier is defined as an RV measurement deviating from both the other measurements and the median posterior model by more than 20 km s$^{-1}$. This threshold is $\sim$3--20 times the typical RV scatter of LRS measurements \citep{Zhang+2026}, the largest among the datasets used in this work.

A second iteration of the fitting is then performed based on the above selection. For sources with $n_{\rm eff} \geq 8$, additional outliers are identified using a $5\sigma$ criterion; no new outliers are flagged for sources with $n_{\rm eff} < 8$ to maximize the retention of RV measurements. The number of RV epochs retained as inliers after this iteration is defined as $n_{\rm used}$ (Table \ref{table2}).

\section{Goodness of Fit Criteria for \emph{Gaia} + RV Fitting}\label{app:score}
\setcounter{figure}{0}
\renewcommand{\thefigure}{B\arabic{figure}}

\begin{sidewaysfigure}[htbp]
    \centering
    \begin{subfigure}[b]{0.48\linewidth}
        \centering
        \includegraphics[width=\linewidth]{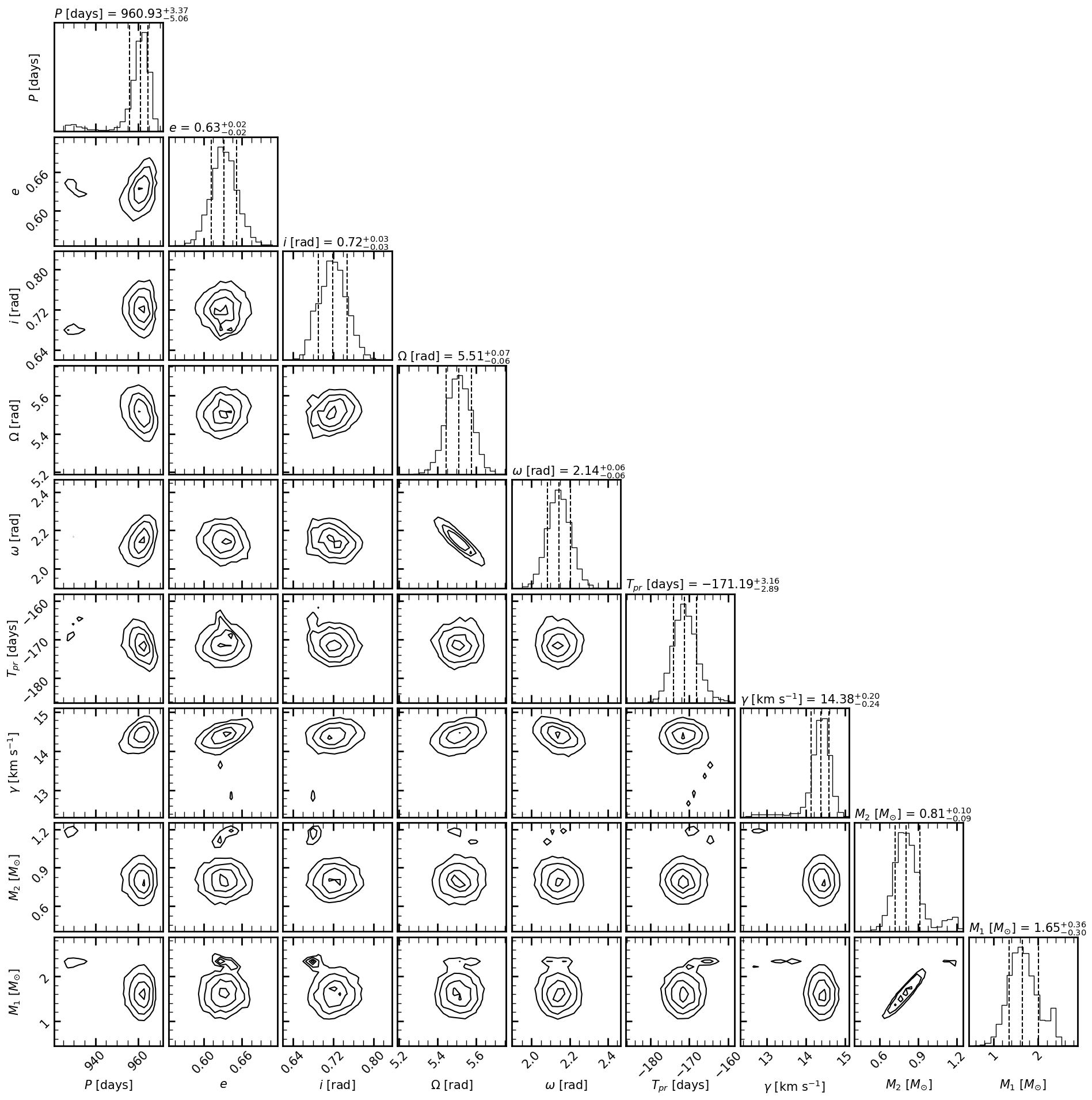}  
        \caption{\emph{Gaia} DR3 231878244865339264 (Good-Fit)}
    \end{subfigure}
    \begin{subfigure}[b]{0.48\linewidth}
        \centering
        \includegraphics[width=\linewidth]{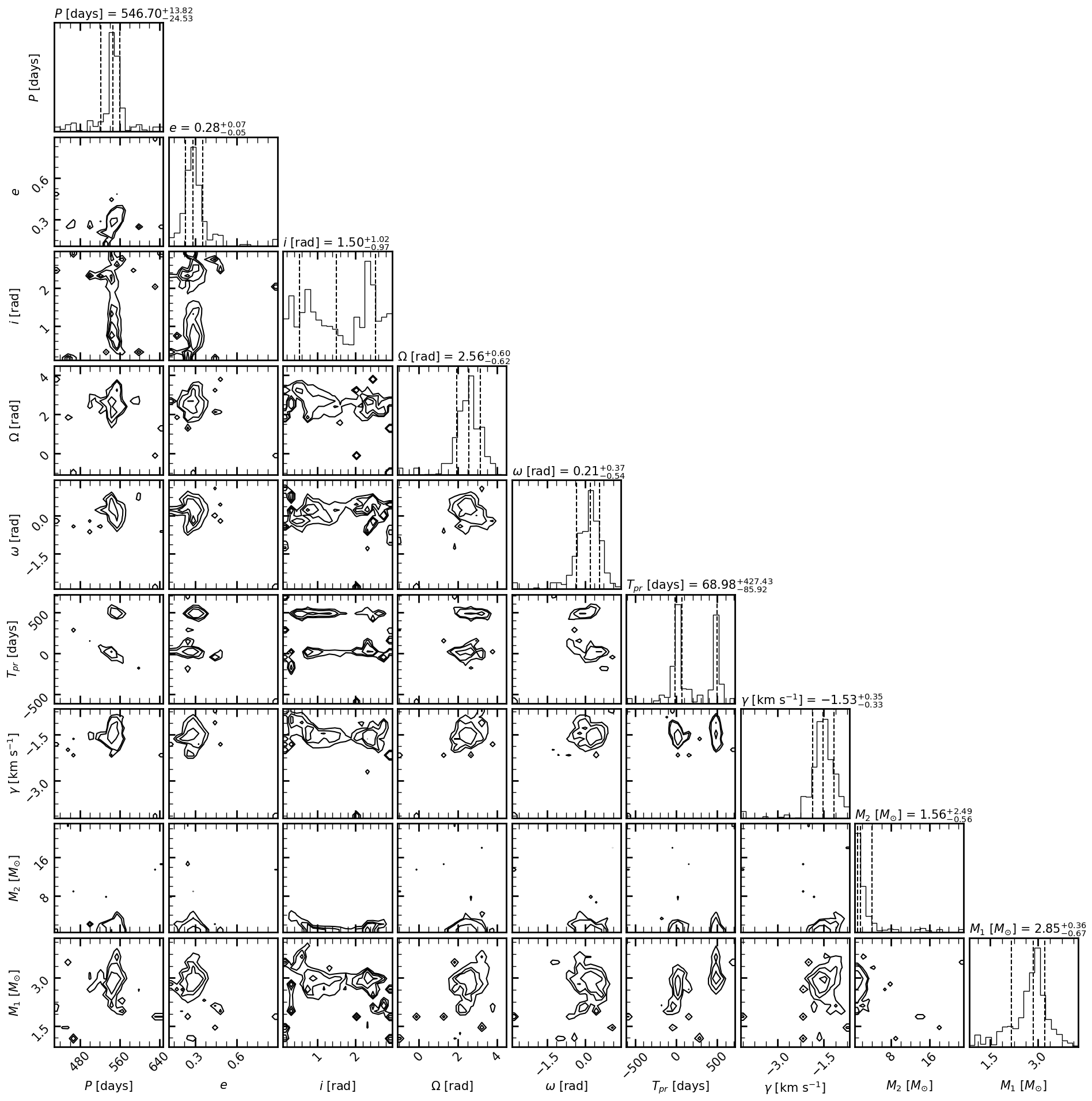} 
        \caption{\emph{Gaia} DR3 157911142723659776 (Poor-Fit)}
    \end{subfigure}
    \caption{Examples of posterior distributions of model parameters for (a) \emph{Gaia} DR3 231878244865339264 (Good-fit) and (b) 157911142723659776 (Poor-fit). $T_{p,r}$ is the periastron time relative to the \emph{Gaia} DR3 reference epoch (MJD = 57388.5). \label{fig:corner}}
\end{sidewaysfigure}

Figure \ref{fig:corner} shows the posterior distributions of the model parameters for two representative sources: \emph{Gaia} DR3 231878244865339264 (Good-fit) and 157911142723659776 (Poor-fit).

Following the method of \citet{Liu+2026}, we evaluate the goodness of fit using three criteria: the reduced chi-square, $\chi^2$, the fraction of RV epochs with residuals within $2\sigma$, $f_{2\sigma}$, and the median absolute normalized residual, $z$:
\begin{equation}
    \chi^2 = \frac{1}{n_{\rm used}}\sum_{i=1}^{n_{\rm used}}\left( \frac{v_{i,\rm obs}-v_{i,\rm mod}}{\sigma_{v,i}} \right)^2,
\end{equation}
\begin{equation}
    f_{2\sigma} = \frac{1}{n_{\rm used}}\sum_{i=1}^{n_{\rm used}}\left(\left|\frac{v_{i,\rm obs}-v_{i,\rm mod}}{\sigma_{v,i}}\right|\leq2\right),
\end{equation}
\begin{equation}
    z={\rm median}\left|\frac{v_{i,\rm obs}-v_{i,\rm mod}}{\sigma_{v,i}}\right|,
\end{equation}
where $v_{i,\rm obs}$ and $v_{i,\rm mod}$ denote the observed and modeled RV values at epoch $i$, respectively, and $\sigma_{v,i}$ is the corresponding measurement uncertainty. A fit is classified as Good if $\chi^{2}\leq 10$, $z\leq 2$, and $f_{2\sigma}\geq 0.6$; otherwise, it is classified as Poor.

\section{Statistical Properties of Good-fit Sources}\label{app:statics}
\setcounter{figure}{0}
\renewcommand{\thefigure}{C\arabic{figure}}

\begin{figure}[htbp]
    \centering
    \begin{subfigure}[b]{0.4\linewidth}
        \centering
        \includegraphics[width=\linewidth]{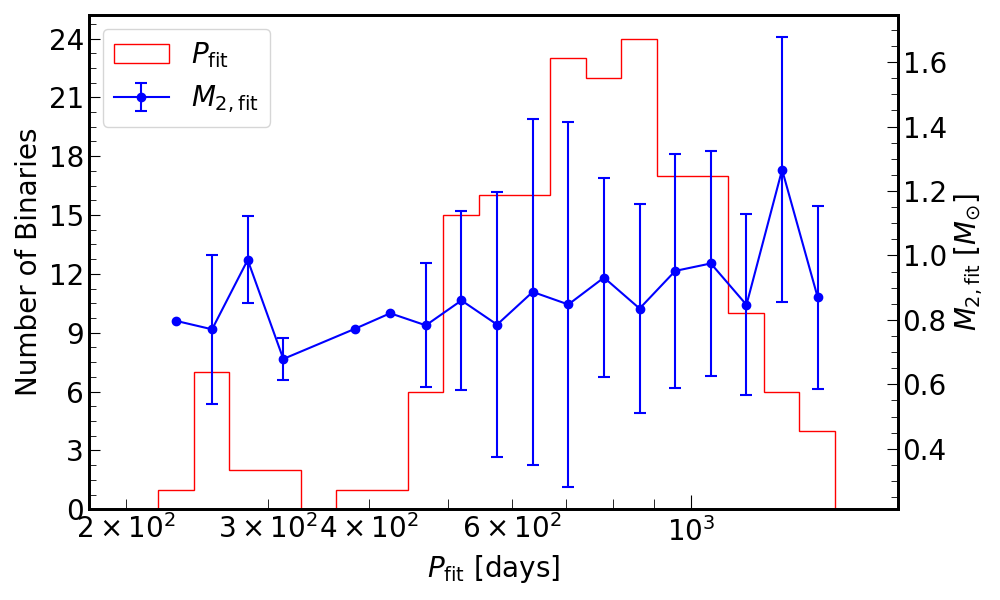}
        \caption{$P_{\rm{fit}}$}
    \end{subfigure}
    \begin{subfigure}[b]{0.4\linewidth}
        \centering
        \includegraphics[width=\linewidth]{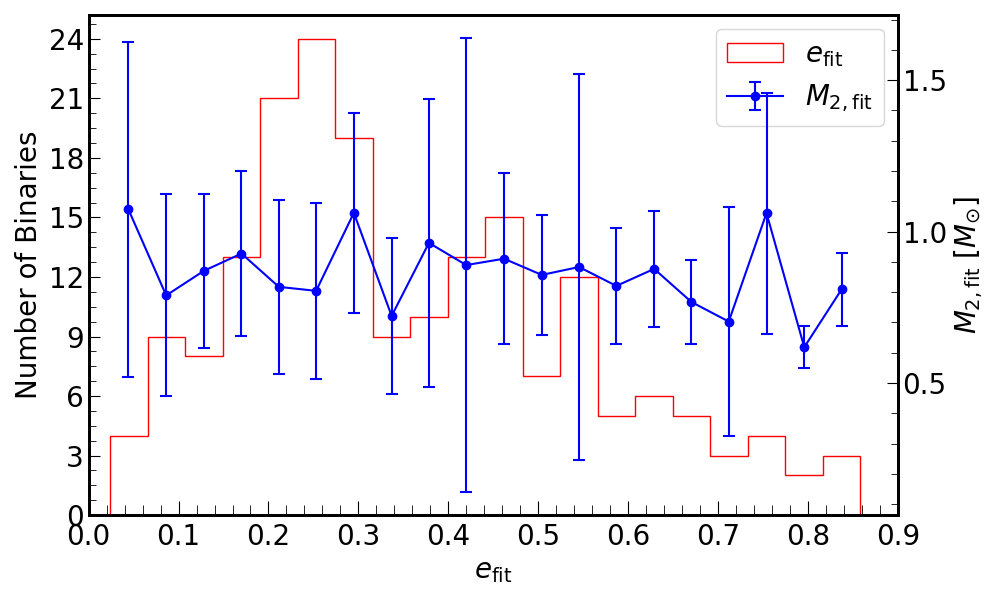}
        \caption{$e_{\rm{fit}}$}
    \end{subfigure}
    \begin{subfigure}[b]{0.4\linewidth}
        \centering
        \includegraphics[width=\linewidth]{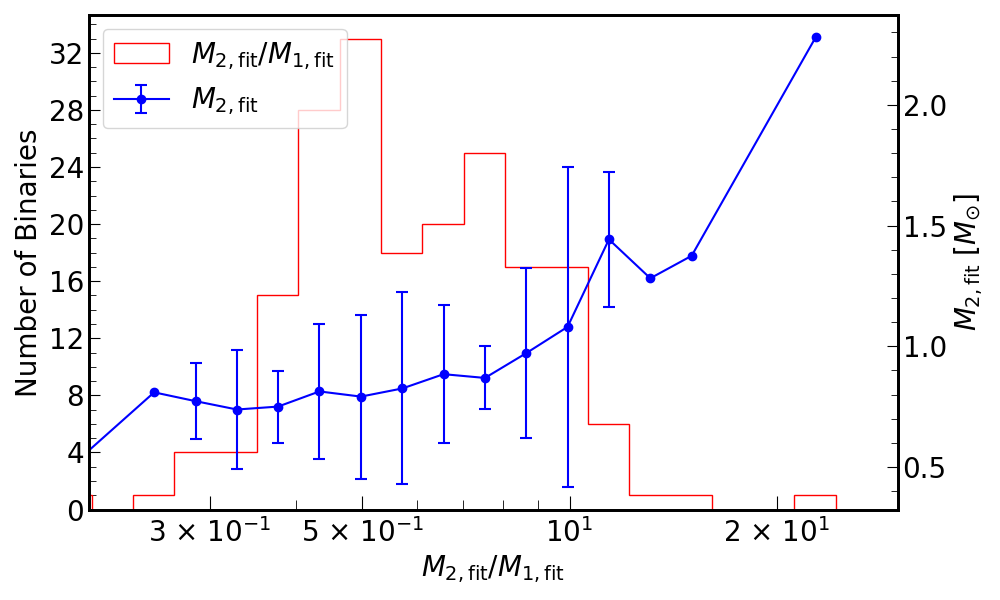}
        \caption{$M_{2,\rm{fit}}/M_{1,\rm{fit}}$}
    \end{subfigure}
    \begin{subfigure}[b]{0.4\linewidth}
        \centering
        \includegraphics[width=\linewidth]{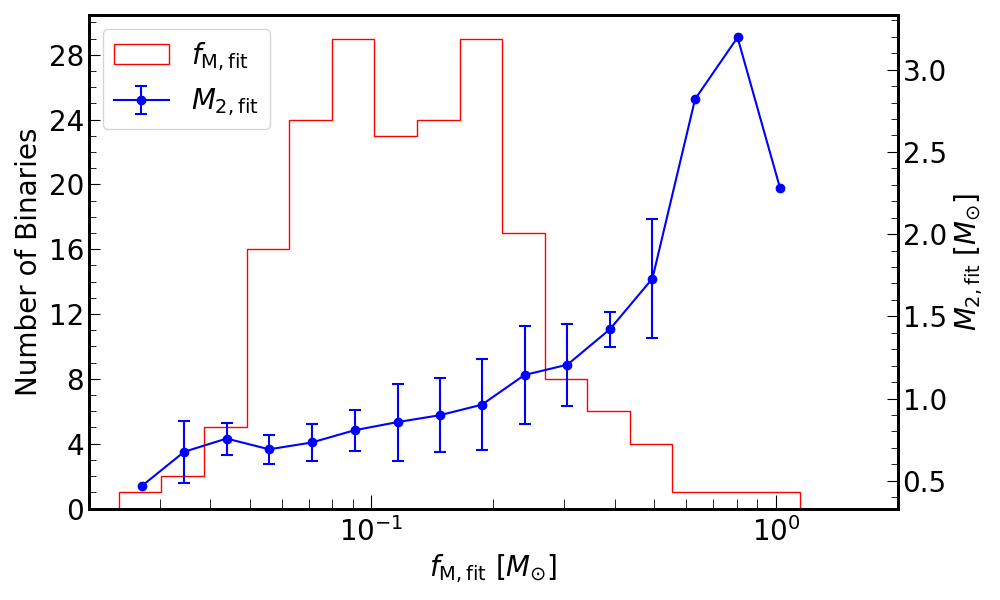}
        \caption{$f_{\rm{M,fit}}$}
    \end{subfigure}
    \caption{
    Distributions of $P_{\rm{fit}}$ (a), $e_{\rm{fit}}$ (b), $M_{2,\rm{fit}}/M_{1,\rm{fit}}$ (c), and $f_{\rm{M,fit}}$ (d), and their correlations with $M_{2,\rm{fit}}$, for the Good-fit sources. Histograms show the parameter distributions, while the blue solid lines with error bars indicate the median and standard deviation of $M_{2,\rm{fit}}$ in each bin; error bars are omitted for bins containing fewer than two sources.\label{fig:M2-Other-parameter}}
\end{figure}

Figure~\ref{fig:M2-Other-parameter} shows the distributions of $P_{\rm{fit}}$, $e_{\rm{fit}}$, $f_{\rm{M,fit}}$, and the mass ratio, $M_{2,\rm{fit}}/M_{1,\rm{fit}}$, for the Good-fit sources, together with their correlations with $M_{2,\rm{fit}}$. The orbital periods are concentrated between $\sim$220 and $\sim$1510 days, with a median of $\sim$750 days. The period distribution also shows a gap of $\sim$90 days centered $\sim$1 yr. The eccentricities span $\sim$0.0--0.9 (median $\sim$0.3), the mass ratios span $\sim$0.2--2.4 (median $\sim$0.6), and the mass functions span $\sim$0.01--1.15 $M_{\odot}$ (median $\sim$0.12 $M_{\odot}$). No clear dependence of $P_{\rm{fit}}$ or $e_{\rm{fit}}$ on $M_{2,\rm{fit}}$ is found. In contrast, both $M_{2,\rm{fit}}/M_{1,\rm{fit}}$ and $f_{\rm{M,fit}}$ exhibit increasing trends with $M_{2,\rm{fit}}$. The apparent correlation between $M_{2,\rm{fit}}/M_{1,\rm{fit}}$ and $M_{2,\rm{fit}}$, however, is mainly driven by the sparsely populated high-mass regime and requires further verification.

\bibliography{sample701_ver2}{}
\bibliographystyle{aasjournalv7}


\end{document}